\documentclass[fp,twocolumn]{jpsj3}
\usepackage{txfonts}
\usepackage{color,graphicx}
\usepackage{dcolumn}
\usepackage{bm}

\def\lesssim{\ \raise.3ex\hbox{$<$}\kern-0.8em\lower.7ex\hbox{$\sim$}\ }
\def\gesim{\ \raise.3ex\hbox{$>$}\kern-0.8em\lower.7ex\hbox{$\sim$}\ }

\title{Single-particle Excitations and Strong Coupling Effects in the BCS-BEC Crossover Regime of a Rare-Earth Fermi Gas with an Orbital Feshbach Resonance}
\author{Soumita Mondal\thanks{soumita\_phy@keio.jp}, Daisuke Inotani, and Yoji Ohashi}
\inst{Department of Physics, Keio University, 3-14-1 Hiyoshi, Yokohama 223-8522, Japan}
\abst{We theoretically investigate normal-state properties of an ultracold Fermi gas with an orbital Feshbach resonance (OFR). Recently, OFR has attracted much attention as a promising pairing mechanism to realize a superfluid $^{173}$Yb Fermi gas.  Including pairing fluctuations within a $T$-matrix approximation, and removing effects of an experimentally inaccessible  deep bound state, we evaluate strong-coupling corrections to single-particle excitations. With increasing the strength of an OFR-induced tunable pairing interaction, the open channel is shown to exhibit the pseudogap phenomenon in the BCS-BEC crossover region, as in the case of a broad magnetic Feshbach resonance (MFR) in $^6$Li and $^{40}$K Fermi gases. We also show that the strong pairing interaction affects the closed channel, leading to the coexistence of particle and hole branches in the single-particle spectral weight. Since the latter phenomenon cannot be observed in the conventional MFR case, it may be viewed as a characteristic strong-coupling phenomenon peculiar to the OFR case.
}
\begin{document}
\maketitle
%%%%%%%%%%%%%%%%%%%%%%%%%%%%%%%%%%%%%%%%%%%%%%%%%%%%%%%%%%%%%%%%%%%%%%%%%%%%%%%
\par
\section{Introduction}
\par
In cold Fermi gas physics, an orbital Feshbach resonance (OFR) has recently attracted much attention as a promising pairing mechanism of a superfluid gas of group 2 (rare earth) Fermi atoms\cite{Zhang,Pagano,Hofer,Cheng,Xu,He,Iskin,Wang,Deng,Soumita,Zou}. The ordinary broad magnetic Feshbach resonance (MFR)\cite{Chin}, which is the pairing mechanism of superfluid $^{40}$K\cite{Jin2004} and $^6$Li Fermi gases\cite{Zwierlein2004,Bartenstein2004,Kinast2004}, strongly relies on the character of the group 1 (alkali metal) elements that one electron occupies the outermost $s$-orbital, giving the total electron spin $S=1/2$. Thus, MFR does not exist in the group 2 elements, because their ground state always has two electrons in the outermost $s$-orbital, giving the total electron spin $S=0$. On the other hand, OFR does {\it not} need any active electron-spin, but only needs two electron orbitals, so that the OFR pairing mechanism is possible in the group 2 elements. OFR has recently been observed in a $^{173}$Yb Fermi gas\cite{Pagano,Hofer}, where the $s$ and $p$ electron orbitals are used.
\par
In a $^{173}$Yb Fermi gas, an {\it optical} Feshbach resonance (OpFR) has been so far discussed as a candidate for the pairing mechanism\cite{Kasa,Blatt,Yan}. However, this scheme is accompanied by a serious short-lifetime problem coming from strong particle loss along with heating. Thus, at present, the prospect of OpFR-mechanism is unclear. On the other hand, although the recently observed OFR\cite{Pagano,Hofer} uses the $^{3}P_0$ electronic {\it excited} state of $^{173}$Yb atom (where one $s$ electron is excited to a $p$-orbital), as well as the $^{1}S_0$ ground state, the dipole transition from the former to the latter is forbidden, so that the lifetime of this excited state is long ($\gesim O(1~{\rm s})$)\cite{Zhang}. Thus, the OFR-mechanism does not suffer from the lifetime problem.
\par
An advantage of Feshbach pairing mechanism is that the resulting interaction strength is tunable, by adjusting the threshold energy of a Feshbach resonance\cite{Chin}. In the conventional MFR case, this advantage has enabled us to realize the so-called BCS (Bardeen-Cooper-Schrieffer)-BEC (Bose-Einstein condensation) crossover in $^{40}$K and $^6$Li Fermi gases\cite{Jin2004,Zwierlein2004,Bartenstein2004,Kinast2004}, where the weak-coupling BCS-type Fermi superfluid continuously changes to the BEC of tightly bound molecules, with increasing the interaction strength\cite{Leggett,NSR,Melo,Randeria1995,Ohashi,Perali,Levin2005,Bloch2008,Giorgini2008}. Since the OFR-induced pairing interaction is also tunable, the achievement of the superfluid phase transition in a $^{173}$Yb Fermi gas with OFR would provide an alternative route to access the BCS-BEC crossover phenomenon.
\par
In this paper, we theoretically investigate single-particle properties of an ultracold Fermi gas with OFR. In particular, we pick up a $^{173}$Yb Fermi gas, because OFR has recently been observed in this rare-earth Fermi gas\cite{Pagano,Hofer}. Since the Fermi degeneracy of a $^{173}$Yb Fermi gas has also been achieved\cite{Fukuhara}, the superfluid phase transition is very promising. In considering this rare-earth Fermi gas, we should note that the observed OFR is relatively narrow (because of the small difference of the nuclear Land\'e $g$-factors between the atomic ground and excited states\cite{Zhang,Boyd}), which required us to deal with both the open channel and closed channels. The resulting system may be viewed as a two band Fermi gas\cite{Walker}, consisting of two atomic states in the open channel and other two atomic states in the closed channel existing above the open channel. We emphasize that this is quite different from the broad MFR case in $^{40}$K and $^6$Li Fermi gases, where the number of atoms in the closed channel is negligibly small\cite{Hulet}, so that one can focus on the open channel. In this case, the open channel is well described by the ordinary BCS model (single-channel model), where effects of the closed channel only remain as the fact that the BCS coupling constant $-U$ is a tunable parameter.
\par
In this paper, including the above-mentioned two-band character, as well as strong pairing fluctuations associated with the OFR-induced pairing interaction within the framework of an $T$-matrix approximation (TMA), we evaluate the single-particle density of states $\rho_{\alpha={\rm o,c}}(\omega)$, as well as the single-particle spectral weight $A_{\alpha={\rm o,c}}({\bm p}, {\omega})$, in the both the open ($\alpha={\rm o}$) and closed ($\alpha={\rm c}$) channels above the superfluid phase transition temperature $T_{\rm c}$. In the open channel, we examine to what extent the BCS-BEC crossover behaviors of these single-particle quantities are similar to the broad MFR case in $^{40}$K and $^6$Li Fermi gases\cite{Tsuchiya,Tsuchiya2}. We also clarify strong coupling corrections to $\rho_{\rm c}(\omega)$ and $A_{\rm c}({\bm p}, \omega)$ in the closed channel. As mentioned previously, the closed channel cannot be examined in alkali metal $^{40}$K and $^6$Li Fermi gases. In this sense, the study of strong-coupling phenomena in the closed channel is an advantage of $^{173}$Yb Fermi gas.
\par
It has theoretically been pointed out\cite{Zhang,He} that, when one simply employs a two-band model to describe a $^{173}$Yb Fermi gas, in addition to a shallow bound state which is responsible to OFR, another deeper bound state is also obtained. Since the latter is nothing to do with OFR, one needs to remove it from the theory, in order to correctly describe the experimental situation. In this paper, we explain this manipulation in the TMA case. We briefly note that we have recently explained this in the case of a Gaussian fluctuation theory\cite{Soumita}.
\par
This paper is organized as follows: In Sec. 2, we present our formulation. Here, we also explain how to remove the unwanted deep bound state from TMA. In Sec. 3, we show our results on the single-particle density of states, as well as the spectral weight, to discuss how these quantities behave in the BCS-BEC crossover region, in both the open and the closed channels. Throughout this paper, we set $\hbar=k_{\rm B}=1$, and the system volume is taken to be unity, for simplicity.
\par
%%%%%%%%%%%%%%%%%%%%%%%%%%%%%%%%%%%%%%%%%%%%%%%%%%%%%%%%%%%%%%%%%%%%%%%%%%%%%%
\begin{figure}[t]
\center
\includegraphics[width=0.45\textwidth]{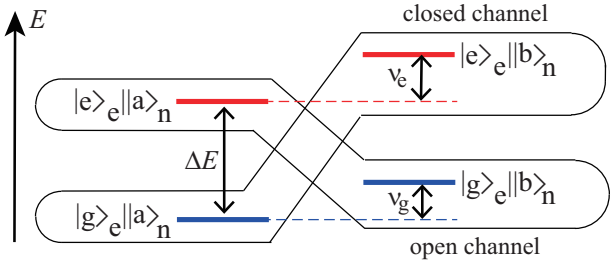}
\caption{(Color online) Schematic energy levels of a model four component Fermi gas. Open channel: $(|{\rm o},\uparrow\rangle,|{\rm o},\downarrow\rangle)=(|{\rm e}\rangle_{\rm e}||{\rm a}\rangle_{\rm n}, |{\rm g}\rangle_{\rm e}||{\rm b}\rangle_{\rm n})$. Closed channel: $(|{\rm c},\uparrow\rangle, |{\rm c},\downarrow\rangle=|{\rm g}\rangle_{\rm e}||{\rm a}\rangle_{\rm n}, |{\rm e}\rangle_{\rm e}||{\rm b}\rangle_{\rm n})$. Here, $|{\rm g}\rangle_{\rm e}$ and $|{\rm e}\rangle_{\rm e}$ are electronic $^{1}S_0$ and $^{3}P_0$ states, respectively. $||{\rm a}\rangle_{\rm n}$ and $||{\rm b}\rangle_{\rm n}$ describe nuclear-spin states. This figure shows the case in the presence of an external magnetic field $B$, giving the Zeeman-energy difference $\nu_{\rm g}$ ($\nu_{\rm e}$) between $|{\rm g}\rangle_{\rm e}||{\rm a}\rangle_{\rm n}$ and $|{\rm g}\rangle_{\rm e}||{\rm b}\rangle_{\rm n}$ ($|{\rm e}\rangle_{\rm e}||{\rm a}\rangle_{\rm n}$ and $|{\rm e}\rangle_{\rm e}||{\rm b}\rangle_{\rm n}$). 
}
\label{Fig1}
\end{figure}
%%%%%%%%%%%%%%%%%%%%%%%%%%%%%%%%%%%%%%%%%%%%%%%%%%%%%%%%%%%%%%%%%%%%%%%%%%%%%%
\par
\section{Formulation}
\par
\subsection{Model two-band Fermi gas with OFR}
\par
To describe a $^{173}$Yb Fermi gas with OFR, we consider a model four component Fermi gas, the energy levels of which are schematically given in Fig. \ref{Fig1}. In this figure, the open channel channel ($|{\rm o},\sigma=\uparrow,\downarrow\rangle$) and the closed channel ($|{\rm c},\sigma=\uparrow,\downarrow\rangle$) consist of, respectively,
\begin{eqnarray}
\left\{
\begin{array}{l}
(|{\rm o},\uparrow\rangle, |{\rm o},\downarrow\rangle)\equiv(|{\rm e}\rangle_{\rm e}||{\rm a}\rangle_{\rm n}, |{\rm g}\rangle_{\rm e}||{\rm b}\rangle_{\rm n}),\\
(|{\rm c},\uparrow\rangle, |{\rm c},\downarrow\rangle)\equiv(|{\rm g}\rangle_{\rm e}||{\rm a}\rangle_{\rm n}, |{\rm e}\rangle_{\rm e}||{\rm b}\rangle_{\rm n}),
\end{array}
\right.
\end{eqnarray}
where $|{\rm g}\rangle_{\rm e}$ and $|{\rm e}\rangle_{\rm e}$ denote the electronic $^{1}S_0$ and $^{3}P_0$ states, respectively. $||{\rm a}\rangle_{\rm n}$ and $||{\rm b}\rangle_{\rm n}$ describe two nuclear-spin states. Under an external magnetic field $B$, the nuclear Zeeman effect brings about the energy difference ($\equiv \nu_{\rm g}$) between $|g\rangle_{\rm e}||{\rm a}\rangle_{\rm n}$ and $|g\rangle_{\rm e}||{\rm b}\rangle_{\rm n}$, as well as the energy difference ($\equiv \nu_{\rm e}$) between $|e\rangle_{\rm e}||{\rm a}\rangle_{\rm n}$ and $|{\rm e}\rangle_{\rm e}||{\rm b}\rangle_{\rm n}$. The magnitude of $\nu_{\rm g}$ is different from that of $\nu_{\rm e}$ due to small difference of nuclear Land\'e $g$-factors between the two cases\cite{Zhang,Pagano,Hofer}. In what follows, we take $\nu_{\rm e}>\nu_{\rm g}$ without loss of generality.
\par
A tunable pairing interaction associated with an orbital Feshbach resonance (OFR) is obtained from an inter-band interaction ($\equiv H_{\rm int}$) between $|{\rm o},\sigma=\uparrow,\downarrow\rangle$ and $|{\rm c},\sigma=\uparrow,\downarrow\rangle$\cite{Zhang,Pagano,Hofer}. Under the assumption that this interaction is independent of nuclear-spins, it is diagonal in terms of the nuclear-spin triplet ($\equiv|+\rangle$) and single ($\equiv|-\rangle$) as\cite{Zhang},
\begin{equation}
H_{\rm int}=U_{++}|+\rangle\langle+|+U_{--}|-\rangle\langle-|.
\label{eq.1}
\end{equation}
Here,
\begin{eqnarray}
|\pm\rangle={1 \over 2}
\left[
|{\rm e},{\rm g}\rangle_{\rm e}\pm |{\rm g},{\rm e}\rangle_{\rm e}
\right]
\left[
||{\rm a},{\rm b}\rangle_{\rm n}
\mp
||{\rm b},{\rm a}\rangle_{\rm n}
\right],
\label{eq.2}
\end{eqnarray}
and $U_{++}$ ($U_{--}$) is an interaction in the nuclear-spin singlet (triplet) channel. In this paper, we treat $U_{\pm\pm}$ as constant values\cite{Zhang,Cheng,He,Soumita}.
\par
Including the interaction $H_{\rm int}$ in Eq. (\ref{eq.1}), as well as the level diagram in Fig. \ref{Fig1}, we consider the model two-band Fermi gas described by the Hamiltonian\cite{note},
\begin{eqnarray}
H
&=&
\sum_{\bm p}
\Bigl[
\bigl[\xi_{\bm p}+\Delta E]
c_{{\rm o},\uparrow,{\bm p}}^\dagger
c_{{\rm o},\uparrow,{\bm p}}
+
\bigl[\xi_{\bm p}+\nu_{\rm g}\bigr]
c_{{\rm o},\downarrow,{\bm p}}^\dagger
c_{{\rm o},\downarrow,{\bm p}}
\Bigr]
\nonumber
\\
&+&
\sum_{\bm p}
\Bigl[
\xi_{\bm p}
c_{{\rm c},\uparrow,{\bm p}}^\dagger
c_{{\rm c},\uparrow,{\bm p}}
+
\bigl[\xi_{\bm p}+\Delta E+\nu_{\rm e}\bigr]
c_{{\rm c},\downarrow,{\bm p}}^\dagger
c_{{\rm c},\downarrow,{\bm p}}
\Bigr]
\nonumber
\\
&+&{1 \over 2}
\sum_{{\bm q},s=\pm}U_{ss}
A_s^{\uparrow\downarrow}({\bm q})^\dagger 
A_s^{\uparrow\downarrow}(-{\bm q}),
\label{eq.3}
\end{eqnarray}
where $c_{\alpha,\sigma,{\bm p}}^\dagger$ is the creation operator of a $^{173}$Yb Fermi atom in the $\alpha~(={\rm o},{\rm c})$ channel, with pseudo-spin $\sigma~(=\uparrow,\downarrow)$. The kinetic energy $\xi_{\bm p}=\varepsilon_{\bm p}-\mu={\bm p}^2/(2m)-\mu$ is measured from the Fermi chemical potential $\mu$, where $m$ is an atomic mass. In Eq. (\ref{eq.3}), $A^{\uparrow\downarrow}_\pm({\bm q})$ is given by
\begin{equation}
A^{\uparrow\downarrow}_\pm({\bm q})=\sum_{\bm p}
\left[
c_{{\rm o},\uparrow,{\bm p}+{\bm q}/2} c_{{\rm o},\downarrow,-{\bm p}+{\bm q}/2}
\pm
c_{{\rm c},\uparrow,{\bm p}+{\bm q}/2} c_{{\rm c},\downarrow,-{\bm p}+{\bm q}/2}
\right].
\label{eq.4}
\end{equation}
\par
In the present case, besides the total number $N=\sum_{{\bm p},\alpha={\rm o},{\rm c},\sigma=\uparrow,\downarrow}c_{\alpha,\sigma,{\bm p}}^\dagger c_{\alpha,\sigma,{\bm p}}$ of Fermi atoms, the number $N_{\rm e}=\sum_{{\bm p},\sigma=\uparrow,\downarrow}c_{{\rm e},\sigma,{\bm p}}^\dagger c_{{\rm e},\sigma,{\bm p}}$ of atoms in the $^{1}S_0$ state ($=|{\rm e}\rangle_{\rm e}$), as well as the number $N_{\rm \uparrow}=\sum_{{\bm p},\alpha={\rm o},{\rm c}}c_{\alpha,\uparrow,{\bm p}}^\dagger c_{\alpha,\uparrow,{\bm p}}$ of atoms in the nuclear $||{\rm a}\rangle_{\rm n}$-spin state, are also conserved. Using these, one may subtract the ``constant" terms $\nu_{\rm g}N/2+[\Delta E+[\nu_{\rm e}-\nu_{\rm g}]/2]N_{\rm e}-[\nu_{\rm e}+\nu_{\rm g}]N_\uparrow/2$ from Eq. (\ref{eq.4})\cite{Zhang}, which gives the two-band Hamiltonian having the form,
\begin{eqnarray}
H
&=&
\sum_{{\bm p},\sigma=\uparrow,\downarrow}
\xi_{\bm p}
c_{{\rm e},\sigma,{\bm p}}^\dagger c_{{\rm e},\sigma,{\bm p}}
\nonumber
\\
&+&
\sum_{{\bm p},\sigma=\uparrow,\downarrow}
\left[\xi_{\bm p}+\nu/2\right]
c_{{\rm c},\sigma,{\bm p}}^\dagger c_{{\rm c},\sigma,{\bm p}}
\nonumber
\\
&+&{U_{\rm intra} \over 2}
\sum_{{\bm p},{\bm p}',{\bm q}}
\left[
c_{{\rm o},\uparrow,{\bm p}'+{\bm q}/2}^\dagger
c_{{\rm o},\downarrow,-{\bm p}'+{\bm q}/2}^\dagger
c_{{\rm o},\downarrow,-{\bm p}+{\bm q}/2}
c_{{\rm o},\uparrow,{\bm p}+{\bm q}/2}
\right.
\nonumber
\\
&{ }&
~~~~~~~~~~~~~~+
\left.
c_{{\rm c},\uparrow,{\bm p}'+{\bm q}/2}^\dagger
c_{{\rm c},\downarrow,-{\bm p}'+{\bm q}/2}^\dagger
c_{{\rm c},\downarrow,-{\bm p}+{\bm q}/2}
c_{{\rm c},\uparrow,{\bm p}+{\bm q}/2}
\right]
\nonumber
\\
&+&{U_{\rm inter} \over 2}
\sum_{{\bm p},{\bm p}',{\bm q}}
\left[
c_{{\rm o},\uparrow,{\bm p}'+{\bm q}/2}^\dagger
c_{{\rm o},\downarrow,-{\bm p}'+{\bm q}/2}^\dagger
c_{{\rm c},\downarrow,-{\bm p}+{\bm q}/2}
c_{{\rm c},\uparrow,{\bm p}+{\bm q}/2}
\right.
\nonumber
\\
&{ }&
~~~~~~~~~~~~~~+
\left.
c_{{\rm c},\uparrow,{\bm p}'+{\bm q}/2}^\dagger
c_{{\rm c},\downarrow,-{\bm p}'+{\bm q}/2}^\dagger
c_{{\rm o},\downarrow,-{\bm p}+{\bm q}/2}
c_{{\rm o},\uparrow,{\bm p}+{\bm q}/2}
\right]
\nonumber
\\
\label{eq.5}
\end{eqnarray}
The first two lines in Eq. (\ref{eq.5}) indicates that the band gap $\nu/2=[\nu_{\rm e}-\nu_{\rm g}]=(2\pi\hbar\times56\Delta_m)B~[{\rm Hz}]$ exists between the open and closed channels (where $\Delta_m=5$ is the difference of the real nuclear-spin quantum number between $||{\rm a}\rangle_{\rm n}$ and $||{\rm b}\rangle_{\rm n}$ for a $^{173}$Yb Fermi gas\cite{Zhang,Xu}). The band gap $\nu/2$ is tunable by an external magnetic field $B$, which is similar to the case of a magnetic Feshbach resonance (MFR), where the energy difference between the open and closed channels are also tuned by an external magnetic field.
\par
In each channel, atoms interact with each other with the coupling constant $U_{\rm intra}=[U_{++}+U_{--}]/2$. Besides this intra-band interaction, the present system also has an inter-band interaction with the coupling constant $U_{\rm inter}=[U_{--}-U_{++}]/2$. From the last line in Eq. (\ref{eq.5}), it may also be viewed as a pair-tunneling between the two channels\cite{Walker}.
\par
%%%%%%%%%%%%%%%%%%%%%%%%%%%%%%%%%%%%%%%%%%%%%%%%%%%%%%%%%%%%%%%%%%%%%%%%%%%%%%
\begin{figure}[t]
\center
\includegraphics[width=0.45\textwidth]{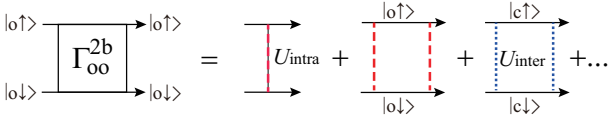}
\caption{(Color online) Two-body scattering matrix $\Gamma_{\rm oo}^{\rm 2b}({\bm q},\omega)$ in the open channel, which is given by summing up all the scattering processes caused by the intra-channel ($U_{\rm intra}$) and inter-channel ($U_{\rm inter}$) interactions. Here, ${\bm q}$ and $\omega$ are the total momentum and total energy of two incident atoms, respectively.}
\label{fig2}
\end{figure}
%%%%%%%%%%%%%%%%%%%%%%%%%%%%%%%%%%%%%%%%%%%%%%%%%%%%%%%%%%%%%%%%%%%%%%%%%%%%%%
\par
To grasp the essence of a tunable pairing interaction associated with OFR, it is convenient to measure the strength of a pairing interaction in the open channel with respect to the $s$-wave scattering $a_s$. This quantity is related to the two-body scattering matrix $\Gamma_{\rm open}^{\rm 2b}({\bm q},\omega)$ in this channel as $4\pi a_s/m=\Gamma_{\rm open}^{\rm 2b}({\bm q}\to {\bm 0},\omega\to 0)$. Summing up the diagrams shown in Fig.\ref{fig2}, one obtains
\begin{equation}
a_s
=a_{\rm intra}
+a_{\rm inter}{\sqrt{m\nu} \over 1-\sqrt{m\nu}a_{\rm intra}}a_{\rm inter}.
\label{eq.6}
\end{equation}
Here, $a_{\rm intra}\equiv[a_++a_-]/2$ and $a_{\rm inter}\equiv [a_--a_+]/2$ are, respectively, the $s$-wave scattering lengths for the intra-band interaction ($U_{\rm intra}$) and the inter-band interaction ($U_{\rm inter}$) when $\nu=0$, where $a_\pm$ are the $s$-wave scattering lengths for $U_{\pm\pm}$, given by
\begin{equation}
{4\pi a_\pm \over m}=
{U_{\pm\pm} \over 
1+U_{\pm\pm}\sum_{\bm p}^{p_{\rm c}}{1 \over 2\varepsilon_{\bm p}}},
\label{eq.7}
\end{equation}
with $p_{\rm c}$ being a high-momentum cutoff. Since the interaction $H_{\rm int}$ is diagonal in the $|\pm\rangle$-basis (see Eq. (\ref{eq.1})), $U_{++}$ and $U_{--}$ do not mix with each other in Eq. (\ref{eq.7}). 
\par
In a $^{173}$Yb Fermi gas, the scattering lengths $a_\pm$ has been measured as $a_+$=1900$a_0$ and $a_-$=200$a_0$\cite{Pagano,Hofer} (where $a_0=0.529~{\rm A}$ is the Bohr radius). It has been shown that these give a shallow two-body bound state with the binding energy $E_+=-1/(ma_+)^2$, as well as a deep bound state with the binding energy $E_=-1/(ma_-)^2\ll E_+$\cite{Zhang,He}. Between the two, the former is responsible for the observed OFR in a $^{173}$Yb Fermi gas\cite{Pagano,Hofer}.
\par
Equation (\ref{eq.6}) shows that, with decreasing the band gap $\nu/2$ near $\nu=1/(ma_{\rm intra}^2)$, the inverse scattering length $a_s^{-1}$ changes its sign, so that the BCS-BEC crossover is expected there. We emphasize that this tunable interaction is not obtained, when the inter-band scattering length $a_{\rm inter}$ is absent. That is, the pair tunneling between the two channels is responsible to this tunable pairing mechanism.
\par
%%%%%%%%%%%%%%%%%%%%%%%%%%%%%%%%%%%%%%%%%%%%%%%%%%%%%%%%%%%%%%%%%%%%%%%%%%%%%%
\begin{figure}[t]
\center
\includegraphics[width=0.3\textwidth]{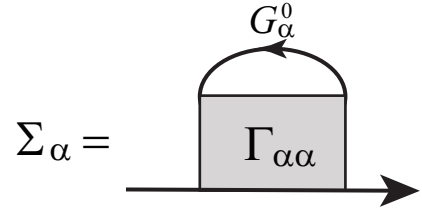}
\caption{(Color online) TMA self-energy $\Sigma_{\alpha={\rm o,c}}$. $G_{\alpha={\rm o,c}}^0$ is the bare single-particle Green's function in Eq. (\ref{eq.10}). The many-body particle-particle scattering matrix $\Gamma_{\alpha\alpha}$ in TMA is diagrammatically given in Fig.\ref{fig4}.}
\label{fig3}
\end{figure}
%%%%%%%%%%%%%%%%%%%%%%%%%%%%%%%%%%%%%%%%%%%%%%%%%%%%%%%%%%%%%%%%%%%%%%%%%%%%%%
\par
%%%%%%%%%%%%%%%%%%%%%%%%%%%%%%%%%%%%%%%%%%%%%%%%%%%%%%%%%%%%%%%%%%%%%%%%%%%%%%
\begin{figure}[t]
\center
\includegraphics[width=0.45\textwidth]{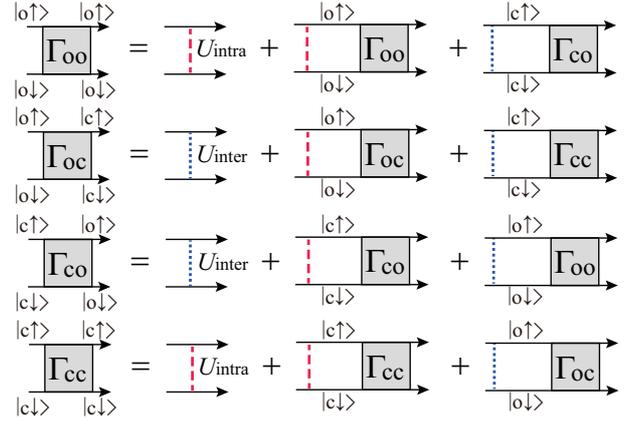}
\caption{(Color online) Particle-particle scattering matrices $\Gamma_{\alpha\alpha'}({\bm q},i\nu_n)$ in TMA ($\alpha,\alpha'={\rm o,c}$). Among them, $\Gamma_{\rm oo}$ and $\Gamma_{\rm cc}$ are used in TMA self-energy in Fig. \ref{fig3}.}
\label{fig4}
\end{figure}
%%%%%%%%%%%%%%%%%%%%%%%%%%%%%%%%%%%%%%%%%%%%%%%%%%%%%%%%%%%%%%%%%%%%%%%%%%%%%%
\par
\subsection{Amended $T$-matrix approximation (ATMA) for a $^{173}$Yb Fermi gas with OFR}
\par
We now include pairing fluctuations within the framework of a $T$-matrix approximation\cite{Perali,Tsuchiya}. In this strong-coupling theory, fluctuation corrections to single-particle excitations are conveniently described by the self-energy $\Sigma_{\alpha={\rm o,c}}({\bm p},i\omega_n)$ in the single-particle thermal Green's function,
\begin{equation}
G_{\alpha={\rm o,c}}({\bm p},i\omega_n)=
{1 \over i\omega_n-\xi_{\bm p}^\alpha-\Sigma_\alpha({\bm p},i\omega_n)}.
\label{eq.8}
\end{equation}
Here, $\xi_{\bm p}^{\rm o}=\varepsilon_{\bm p}-\mu$ and $\xi_{\bm p}^{\rm c}=\varepsilon_{\bm p}+\nu/2-\mu$ are the kinetic energy in the open and closed channels, respectively. $\omega_n$ is the fermion Matsubara frequency. The TMA self-energy $\Sigma_\alpha({\bm p},i\omega_n)$ is diagrammatically described as Fig. \ref{fig3}, which gives,
\begin{equation}
\Sigma_{\alpha}({\bm p},i\omega_n)=
T\sum_{{\bm q},\nu_n}\Gamma_{\alpha\alpha}({\bm q},i\nu_n)
G_\alpha^0({\bm q}-{\bm p},i\nu_n-i\omega_n),
\label{eq.9}
\end{equation}
where $\nu_n$ is the boson Matsubara frequency, and 
\begin{equation}
G_{\alpha={\rm o,c}}^0({\bm p},i\omega_n)=
{1 \over i\omega_n-\xi_{\bm p}^\alpha}
\label{eq.10}
\end{equation}
is the bare single-particle thermal Green's function in the $\alpha$-channel. 
\par
When one extends TMA developed in the single-channel BCS model\cite{Perali,Tsuchiya,Tsuchiya2} to the present two-channel case, the particle-particle scattering matrix $\Gamma_{\alpha\alpha}({\bm q},i\nu_n)$ in Eq. (\ref{eq.9}) is obtained by summing up the diagrams in Fig. \ref{fig4}. The result is
\begin{eqnarray}
&{}&
\hskip-9mm\left(
\begin{array}{cc}
\Gamma_{\rm oo}({\bm q},i\nu_n)& 
\Gamma_{\rm oc}({\bm q},i\nu_n)\\
\Gamma_{\rm co}({\bm q},i\nu_n)& 
\Gamma_{\rm cc}({\bm q},i\nu_n)\\
\end{array}
\right)
\nonumber
\\
&=&
\left[
1
-
\left(
\begin{array}{cc}
U_{\rm intra} U_{\rm inter}\\
U_{\rm inter} U_{\rm intra}
\end{array}
\right)
\left(
\begin{array}{cc}
\Pi_{\rm o}({\bm q},i\nu_n)& 0\\
0& \Pi_{\rm c}({\bm q},i\nu_n)
\end{array}
\right)
\right]^{-1}
\nonumber
\\
&\times&
\left(
\begin{array}{cc}
U_{\rm intra} U_{\rm inter}\\
U_{\rm inter} U_{\rm intra}
\end{array}
\right).
\label{eq.11}
\end{eqnarray}
Here,
\begin{eqnarray}
\Pi_{\alpha={\rm o,c}}(\bm q,i\nu_n)
= 
\sum_{\bm p}
{
1-f(\xi_{{\bm p}+{\bm q}/2}^\alpha)-f(\xi_{-{\bm p}+{\bm q}/2}^\alpha)
\over
i\nu_n-\xi_{{\bm p}+{\bm q}/2}^\alpha-\xi_{-{\bm p}+{\bm q}/2}^\alpha
}
\label{eq.12}
\end{eqnarray}
is the lowest-order pair correlation function in the $\alpha$-channel, where $f(x)$ is the Fermi distribution function. 
\par
However, $\Gamma_{\alpha\alpha'}({\bm q},i\nu_n)$ in Eq. (\ref{eq.11}) is known to involve contribution from, not only the experimentally accessible bound state with shallow binding energy $E_+^{\rm bind}$= -1/(m$a_+^2)$, but also another bound state which is experimentally inaccessible because of the very deep energy level $E_-^{\rm bind}=-1/(ma_-^2)\ll E_+$\cite{Zhang,He,Soumita} (As mentioned previously, $a_+$=1900$a_0$ and $a_-$=200$a_0$ in a $^{173}$Yb Fermi gas\cite{Pagano,Hofer}.) To correctly describe the recent experimental situation for a $^{173}$Yb Fermi gas\cite{Pagano,Hofer}, one needs to remove the latter contribution from the theory. For this purpose, we diagonalize the $2\times 2$-matrix ${\hat \Gamma}=\{\Gamma_{\alpha\alpha'}\}$ as
\small
\begin{equation}
{\hat \Gamma}_{\rm D}\equiv {\hat W}{\hat \Gamma}{\hat W}^{-1}=
\left(
\begin{array}{cc}
{\lambda}_+(\bm q,i\nu_n)&0 \\
0&{\lambda}_-(\bm q,i\nu_n)
\end{array}
\right),
\label{eq.13}
\end{equation}
where 
\begin{equation}
\hat{W}=
\left(
\begin{array}{cc}
{X-\sqrt{Y} \over 2U_{\rm inter}}&
{X+\sqrt{Y} \over 2U_{\rm inter}}\\
1&1
\end{array}
\right).
\label{eq.14}
\end{equation}
In Eq. (\ref{eq.14}), $X=[\Pi_{\rm o}-\Pi_{\rm c}][U_{\rm intra}^2-U_{\rm inter}^2]$, and $Y=[\Pi_{\rm o}-\Pi_{\rm c}]^2[U_{\rm intra}^2-U_{\rm inter}^2]^2+4U_{\rm inter}^2$. The eigen-values $\lambda_\pm({\bm q},i\nu_n)$ in Eq. (\ref{eq.13}) are given by
\begin{eqnarray}
\lambda_\pm
={1 \over 2}
{
[U_{\rm inter}^2-U_{\rm intra}^2]
[\Pi_{\rm o}+\Pi_{\rm c}]+2U_{\rm intra}
\pm
\sqrt{Y}
\over
1-U_{\rm intra}[\Pi_{\rm o}+\Pi_{\rm c}]
+[U_{\rm intra}^2-U_{\rm inter}^2]\Pi_{\rm o}\Pi_{\rm c}
}.
\label{eq.15}
\end{eqnarray}
To see which corresponds to the shallow bound state responsible for OFR, it is convenient to take the two particle limit at $T=0$ at the vanishing band gap $\nu=0$. In this extreme case, setting $\mu=0$ and $f(x)=0$ in the pair-correlation function $\Pi_\alpha$ in Eq. (\ref{eq.12}), one finds that the analytic continued $\lambda_\pm({\bm q}=0,i\nu_n\to\omega+i\delta)$ becomes
\begin{equation}
\lambda_\pm({\bm q}=0,\omega)
=
{
\displaystyle
{4\pi a_\pm \over m}
\over 
\displaystyle
1+{4\pi a_\pm \over m}\sum_{\bm p}
\left[
{1 \over 2\varepsilon_{\bm p}-\omega}-{1 \over 2\varepsilon_{\bm p}}
\right]
},
\label{eq.16}
\end{equation}
where the scattering length $a_\pm$ is given in Eq. (\ref{eq.7}). Noting that the condition for the pole of Eq. (\ref{eq.16}) is just the same as the equation for the two-body bound state, we find that $\lambda_\pm({\bm q}=0,\omega)$ diverges at the bound state energy $\omega=E_\pm=-1/(ma_\pm^2)$. That is, $\lambda_+$ and $\lambda_-$ correspond to the shallow and deep bound states, respectively. To conclude, to describe the $^{173}$Yb case, one should replace the particle-particle scattering matrix ${\hat \Gamma}({\bm q},i\nu_n)$ by
\begin{eqnarray}
{\hat {\tilde \Gamma}}=
\left(
\begin{array}{cc}
{\tilde \Gamma}_{\rm oo}&
{\tilde \Gamma}_{\rm oc}\\
{\tilde \Gamma}_{\rm co}&
{\tilde \Gamma}_{\rm cc}
\end{array}
\right)
=
{\hat W}
\left(
\begin{array}{cc}
\lambda_+({\bm q},i\nu_n)&
0\\
0&
0
\end{array}
\right)
{\hat W}^{-1}
\label{eq.17}
\end{eqnarray}
The resulting {\it amended} TMA (ATMA) for a $^{173}$Yb Fermi gas uses the following self-energy: 
\begin{equation}
{\tilde \Sigma}_{\alpha}({\bm p},i\omega_n)=
T\sum_{{\bm q},\nu_n}{\tilde \Gamma}_{\alpha\alpha}({\bm q},i\nu_n)
G_\alpha^0({\bm q}-{\bm p},i\nu_n-i\omega_n).
\label{eq.18}
\end{equation}
\par
As usual, we determine the superfluid phase transition temperature $T_{\rm c}$ from the Thouless criterion\cite{Thouless}, stating that the superfluid instability occurs when the particle-particle scattering matrix has a pole in the low-energy and low-momentum limit. In ATMA, the poles of both ${\tilde \Gamma}_{\rm oo}({\bm q}=0,i\nu_n=0)$ and ${\tilde \Gamma}_{\rm cc}({\bm q}=0,i\nu_n=0)$ are commonly determined from $\lambda_+({\bm q}=0,i\nu_n=0)^{-1}=0$, which gives the $T_{\rm c}$ equation,
\begin{equation}
1-U_{\rm intra}[\Pi_{\rm o}(0,0)+\Pi_{\rm c}(0,0)]+[U_{\rm intra}^2-U_{\rm inter}^2]\Pi_{\rm o}(0,0)\Pi_{\rm c}(0,0)=0.
\label{eq.19}
\end{equation}
Below $T_{\rm c}$, both the open and closed channels are in the superfluid phase. To see how OFR works in Eq. (\ref{eq.19}), it is convenient to rewrite this equation into the form being similar to the ordinary BCS gap equation at $T_{\rm c}$ as 
\begin{equation}
1=-{4\pi {\tilde a}_s \over m}
\left[
{1 \over 2\xi_{\rm p}^{\rm o}}\tanh{\xi_{\rm p}^{\rm o} \over 2T}
-
{1 \over 2\varepsilon_{\bm p}}
\right],
\label{eq.20}
\end{equation}
Here, the effective scattering length ${\tilde a}_s$ is given by
\begin{equation}
{\tilde a}_s=a_{\rm intra}
+
a_{\rm inter}
{\displaystyle {4\pi \over m}{\tilde \Pi}_{\rm c}(0,0) 
\over 
\displaystyle
1-{4\pi a_{\rm intra} \over m}{\tilde \Pi}_{\rm c}(0,0)}
a_{\rm inter},
\label{eq.21}
\end{equation}
where
\begin{equation}
{\tilde \Pi}_{\rm c}(0,0)=-\sum_{\bm p}
\left[
{1 \over 2\xi_{\bm p}^{\rm c}}\tanh{\xi_{\bm p}^{\rm c} \over 2T}
-
{1 \over 2\varepsilon_{\rm p}}
\right].
\label{eq.22}
\end{equation}
Comparing ${\tilde a}_s$ with Eq. (\ref{eq.6}), we find that the second term in Eq. (\ref{eq.21}) is the OFR-induced tunable interaction extended to the many-particle case. Indeed, in the low density limit ($\mu\to 0$) at $T=0$, Eq. (\ref{eq.22}) becomes $m\sqrt{m\nu}/(4\pi)$, so that Eq. (\ref{eq.22}) is reduced to the two-body scattering length $a_s$ in Eq. (\ref{eq.6}). 
\par
\par
%%%%%%%%%%%%%%%%%%%%%%%%%%%%%%%%%%%%%%%%%%%%%%%%%%%%%%%%%%%%%%%%%%%%%%%%%%%%%
\begin{figure}[t]
\center
\includegraphics[width=0.35\textwidth]{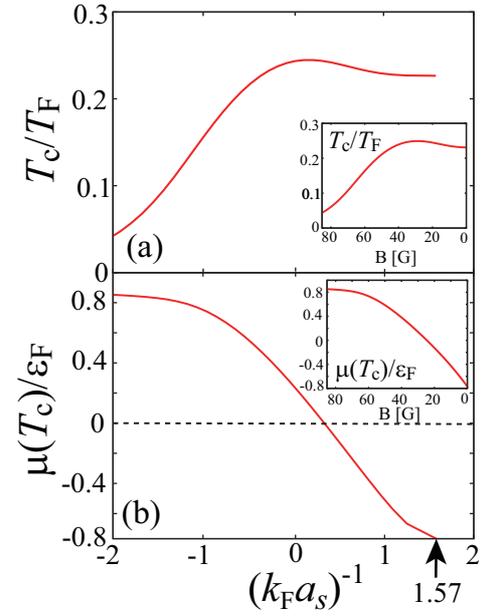}
\caption{(Color online) Self-consistent solutions for the superfluid phase transition temperature $T_{\rm c}$ (a), as well as the chemical potential $\mu(T_{\rm c})$ (b), in the BCS-BEC crossover region. The interaction strength is measured in terms of the inverse scattering length $a_s$ in the open channel (where $k_{\rm F}$ is the Fermi momentum). Since we are setting $\hbar=k_{\rm B}=1$, The Fermi temperature $T_{\rm F}$ equals the Fermi energy $\varepsilon_{\rm F}=k_{\rm F}^2/(2m)$. For experimental convenience, we also plot the magnetic field ($B$) dependence of $T_{\rm c}$ and $\mu(T_{\rm c})$ in the insets in the upper and lower panels, respectively.}
\label{fig5}
\end{figure}
%%%%%%%%%%%%%%%%%%%%%%%%%%%%%%%%%%%%%%%%%%%%%%%%%%%%%%%%%%%%%%%%%%%%%%%%%%%%%%
\par
We numerically solve the coupled $T_{\rm c}$-equation (\ref{eq.20}) with the equation for the total number $N=N_{\rm o}+N_{\rm c}$ of Fermi atoms, to self-consistently determine $T_{\rm c}$ and $\mu(T_{\rm c})$. Here, the particle number $N_{\alpha={\rm o,c}}$ in the $\alpha$-channel is calculated from the ATMA single-particle Green's function in Eq. (\ref{eq.8}) as
\begin{equation}
N_\alpha=2T\sum_{{\bm p},\omega_n}G_\alpha({\bm p},i\omega_n).
\label{eq.23}
\end{equation}
In the normal state above $T_{\rm c}$, we only deal with the number equation $N$, to determine $\mu(T>T_{\rm c})$.
\par
In numerical calculations, we always assume that $N_{\alpha,\uparrow}=N_{\alpha,\downarrow}$, where $N_{\alpha={\rm o,c},\sigma}$ is the number of Fermi atoms in the $|\alpha,\sigma\rangle$-state. For the interaction parameters, we take the previously mentioned experimental values $a_+=1900a_0$ and $a_-=200a_0$\cite{Pagano,Hofer}, giving $a_{\rm intra}=[a_++a_-]/2=1050a_0$ and $a_{\rm inter}=[a_--a_+]/2=-850a_0$. We also take the particle density $n=5\times 10^{13}~{\rm cm}^{-3}$ observed in the trap center of a $^{173}$Yb Fermi gas\cite{Pagano}. Using the Fermi momentum $k_{\rm F}=[3\pi^2 n]^{1/3}$ in the case when all the Fermi atoms occupy the open channel, we measure the interaction strength in terms of $(k_{\rm F}a_s)^{-1}$ (where $a_s$ is give in Eq. (\ref{eq.6})). In this case, the vanishing band gap between the two channels ($\nu=0$, or $B=0$) occurs when $(k_{\rm F}a_s)^{-1}\simeq 1.57$. (Note that the relation between $\nu$ and an external magnetic field $B$ is given below Eq. (\ref{eq.5}).) Since we assume $\nu\ge 0$, we consider the region $(k_{\rm F}a_s)^{-1}\le 1.57$ in this paper. For later convenience, we introduce the Fermi energy $\varepsilon_{\rm F}$, as well as the Fermi temperature $T_{\rm F}$, for the assumed Fermi momentum $k_{\rm F}$ as $\varepsilon_{\rm F}=T_{\rm F}=k_{\rm F}^2/(2m)$.
\par
\begin{figure}[t]
\center
\includegraphics[width=0.35\textwidth]{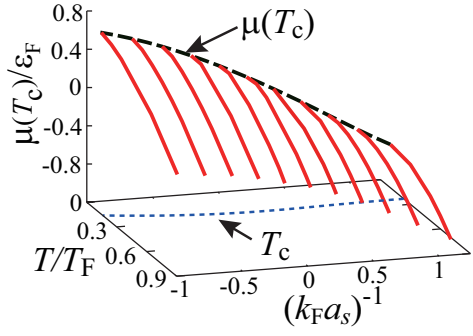}
\caption{(Color online) Self-consistent solution for the Fermi chemical potential $\mu(T)$ above $T_{\rm c}$.}
\label{fig6}
\end{figure}

Before ending section, we summarize in Figs. \ref{fig5} and \ref{fig6} the self-consistent solutions for the superfluid phase transition temperature $T_{\rm c}$, as well as the Fermi chemical potential $\mu(T\ge T_{\rm c})$. These will be used in calculating single-particle excitations in Sec. 3. We briefly note that the interaction dependence of $T_{\rm c}$ and that of $\mu(T_{\rm c})$ shown in Fig. \ref{fig5} are qualitatively the same as the BCS-BEC crossover behaviors of these quantities known in the single-channel case with a broad MFR\cite{Perali,Tsuchiya,Tsuchiya2}. We also note that similar results to Figs. \ref{fig5} and \ref{fig6} have been obtained\cite{Xu,Soumita} within the framework of the strong-coupling theory developed by Nozi\`eres and Schmitt-Rink (NSR)\cite{NSR}.
\par
%%%%%%%%%%%%%%%%%%%%%%%%%%%%%%%%%%%%%%%%%%%%%%%%%%%%%%%%%%%%%%%%%%%%%%%%%%%%%%
%%%%%%%%%%%%%%%%%%%%%%%%%%%%%%%%%%%%%%%%%%%%%%%%%%%%%%%%%%%%%%%%%%%%%%%%%%%%%
\par
%%%%%%%%%%%%%%%%%%%%%%%%%%%%%%%%%%%%%%%%%%%%%%%%%%%%%%%%%%%%%%%%%%%%%%%%%%%%%
\begin{figure}[t]
\center
\includegraphics[width=0.48\textwidth]{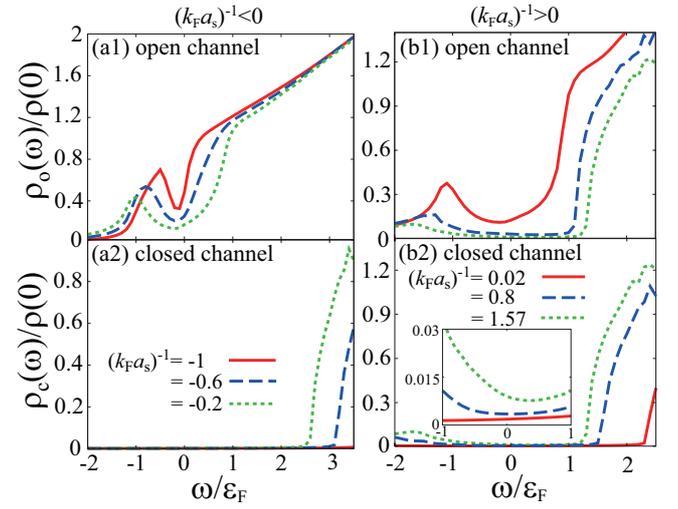}
\caption{(Color online) Calculated single-particle density of states (DOS) $\rho(\omega)$ in the BCS-BEC crossover region at $T_{\rm c}$. (a1) and (b1) show the results $\rho_{\rm o}(\omega)$ in the open channel. (a2) and (b2) show the results $\rho_{\rm c}(\omega)$ in the closed channel. $\rho(0)=mk_{\rm F}/(2\pi^2)$ is DOS of a free Fermi gas at the Fermi level. The inset in panel (b2) shows $\rho_{\rm c}(\omega)$ magnified around $\omega=0$. In panel (a2), $\rho_{\rm c}(\omega)$ at $(k_{\rm F}a_s)^{-1}=-1$ only becomes large when $\omega/\varepsilon_{\rm F}\gesim 3.5$, so that this weak-coupling result almost vanishes within the $x$-range $\omega/\varepsilon_{\rm F}=[0,3.5]$ shown in this figure.}
\label{fig7}
\end{figure}
%%%%%%%%%%%%%%%%%%%%%%%%%%%%%%%%%%%%%%%%%%%%%%%%%%%%%%%%%%%%%%%%%%%%%%%%%%%%%
\par
\section{Single-particle excitations in the BCS-BEC crossover regime of a $^{173}$Yb Fermi gas}
\par
Figure \ref{fig7} shows the single-particle density of states (DOS) $\rho_{\alpha={\rm o,c}}(\omega)$ at $T_{\rm c}$, given by
\begin{equation}
\rho_\alpha(\omega)=\sum_{\bm p}A_\alpha({\bm p},\omega).
\label{eq.24}
\end{equation}
Here, the single-particle spectral weight $A_\alpha({\bm p},\omega)$ is related to the analytic continued ATMA single-particle Green's function in Eq. (\ref{eq.8}) as
\begin{equation}
A_\alpha({\bm p},\omega)=-{1 \over \pi}
{\rm Im}
\left[
G_\alpha({\bm p},i\omega_n\to\omega+i\delta)
\right],
\label{eq.25}
\end{equation}
where $\delta$ is an infinitesimally small positive number. In this paper, we carry out the analytic continuation by the Pad\'e approximation\cite{Vidberg}. In the open channel (panels (a1) and (b1)), one sees the so-called pseudogap phenomenon as in a Fermi gas with a broad MFR\cite{Perali,Tsuchiya,Tsuchiya2,Chen,Watanabe,Miki}. That is, $\rho_{\rm o}(\omega)$ exhibits a dip structure around $\omega=0$ in the BCS regime when $(k_{\rm F}a_s)^{-1}=-1$, in spite of the fact that the superfluid order parameter $\Delta$ vanishes at $T_{\rm c}$. This pseudogap structure grows as one passes through the BCS-BEC crossover region, as seen in Figs. \ref{fig7}(a1) and (b1). 
\par
Figure \ref{fig7}(b1) shows that $\rho_{\rm o}(\omega)$ at the strongest interaction $(k_{\rm F}a_s)^{-1}=1.57$ (or $\nu=0$) already has the almost fully gapped structure expected deep inside the BEC regime. Thus, although the present OFR case cannot go beyond this interaction strength because of the restriction $\nu\ge 0$, we can practically investigate the BCS-BEC crossover behavior of DOS in the open channel, even under this restriction.
\par
\begin{figure}[t]
\center
\includegraphics[width=0.35\textwidth]{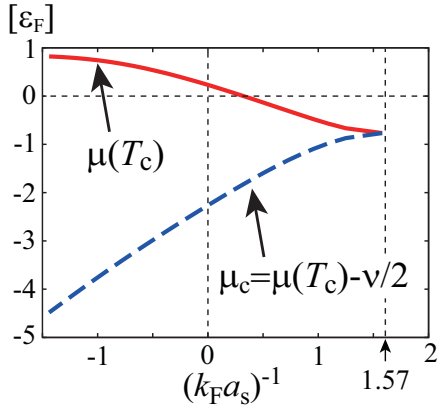}
\caption{(Color online) Effective chemical potential $\mu_{\rm c}=\mu-\nu/2$ in the closed channel at $T_{\rm c}$. Note that the single-particle dispersion in the close channel is given by $\xi_{\bm p}^{\rm c}=\varepsilon_{\bm p}-\mu_{\rm c}$. For comparison, we also plot $\mu(T_{\rm c})$.}
\label{fig8}
\end{figure}

On the other hand, DOS $\rho_{\rm c}(\omega)$ in the closed channel exhibits  different interaction dependence from $\rho_{\rm o}(\omega)$, as shown in Figs. \ref{fig7}(a2) and (b2): In the BCS side shown in panel (a2) ($(k_{\rm F}a_s)^{-1}<0$), one does not see any remarkable strong-coupling corrections to $\rho_{\rm c}(\omega)$, in contrast to the pseudo-gapped $\rho_{\rm o}(\omega)$ in the open channel. Large excitation threshold energy ($\equiv E_{\rm th}$) is only seen (above which $\rho_{\rm c}(\omega)$ becomes large), reflecting the large band gap $\nu/2$ between the open and closed channels. As shown in Fig. \ref{fig8}, the effective chemical potential in the closed channel, defined by
\begin{equation}
\mu_{\rm c}(T) \equiv \mu(T)-\nu/2,
\label{eq.eff}
\end{equation}
is always {\it negative} at $T_{\rm c}$, even in the weak-coupling BCS regime (where $\mu(T_{\rm c})$ is positive). Retaining this and ignoring any other effects, one obtains non-zero value of DOS in the closed channel, only when $\omega\ge |\mu-\nu/2|$. At $(k_{\rm F}a_s)^{-1}=-0.6$, the excitations threshold $E_{\rm th}\simeq 3\varepsilon_{\rm F}$ seen in Fig. \ref{fig7}(a2) (where $\varepsilon_{\rm F}=k_{\rm F}^2/(2m)$) agrees with $\mu_{\rm c}=\mu(T_{\rm c})-\nu/2=-3.1\varepsilon_{\rm F}$ at this interaction strength (see Fig. \ref{fig8}). 
\par
%%%%%%%%%%%%%%%%%%%%%%%%%%%%%%%%%%%%%%%%%%%%%%%%%%%%%%%%%%%%%%%%%%%%%%%%%%%%%%
%%%%%%%%%%%%%%%%%%%%%%%%%%%%%%%%%%%%%%%%%%%%%%%%%%%%%%%%%%%%%%%%%%%%%%%%%%%%%
\par
The interaction becomes strong with decreasing $\nu/2$ (see Eq. (\ref{eq.6})). As a result, the excitation threshold $E_{\rm th}~(\sim |\mu-\nu/2|)$ become small with increasing the interaction strength, as seen in Fig. \ref{fig7}(a2). With further increasing the interaction strength to enter the BEC side ($k_{\rm F}a_s)^{-1}>0$), we find in Fig. \ref{fig7}(b2) that $\rho_{\rm c}(\omega)$ below $E_{\rm th}$ gradually increases. When the band gap $\nu/2$ vanishes at $(k_{\rm F}a_s)^{-1}=1.57$, $\rho_{\rm c}(\omega)$ coincides with $\rho_{\rm o}(\omega)$, as expected. To conclude, DOS $\rho_{\rm c}(\omega)$ in the closed channel continuously changes from the band gap structure (with large excitation threshold $E_{\rm th}$) to the molecular gapped structure (with the energy gap associated with the binding energy of a two-body bound state).
\par
\begin{figure}[t]
\center
\includegraphics[width=0.4\textwidth]{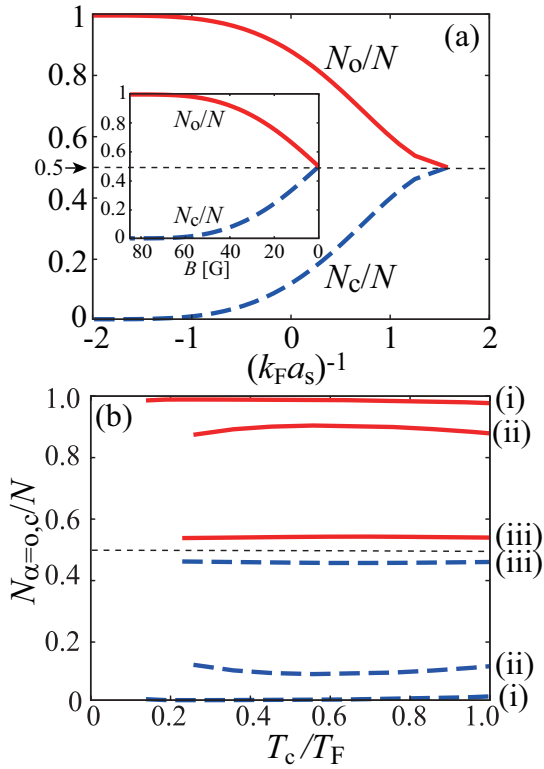}
\caption{(Color online) (a) Calculated number $N_{\rm o}$ ($N_{\rm c}$) of atoms in the open (closed) channel at $T_{\rm c}$. The inset shows the magnetic field dependence of $N_{\alpha={\rm o},{\rm c}}$. (b) $N_{\rm o}$  (upper three lines) and $N_{\rm c}$ (lower three lines) as functions of temperature. (i) $(k_{\rm F}a_s)^{-1}=-1$. (ii) $(k_{\rm F}a_s)^{-1}=0.02$. $(k_{\rm F}a_s)^{-1}=1.24$.}
\label{fig9}
\end{figure}

Because the number equation (\ref{eq.23}) can be written as
\begin{equation}
N_\alpha=\int_{-\infty}^\infty d\omega f(\omega)\rho_\alpha(\omega),
\label{eq.26}
\end{equation}
the large gap structure of $\rho_{\rm c}(\omega)$ seen in Fig. \ref{fig7}(a2)  makes us expect that the number $N_{\rm c}$ of atoms in the closed channel almost vanishes in the BCS side. However, Fig. \ref{fig9}(a) shows that $N_{\rm c}/N$ actually amounts to about 0.1 around the unitarity limit ($(k_{\rm F}a_s)^{-1}=0$), implying that $\rho_{\rm c}(\omega)$ is very small but is still non-zero even below the excitation threshold $E_{\rm th}$ in the BCS side. 
\par
\begin{figure}[t]
\center
\includegraphics[width=0.45\textwidth]{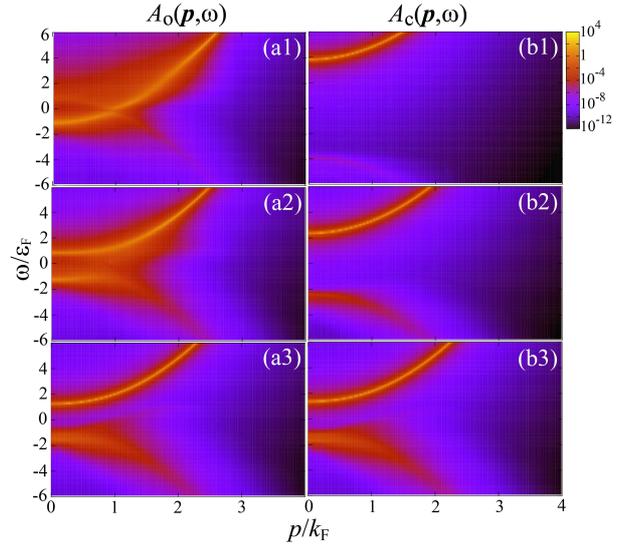}
\caption{(Color online) Calculated intensity of single-particle spectral weight $A_{\alpha={\rm o,c}}({\bm p},\omega)$ at $T_{\rm c}$. The left and right panels show the results in the open and close channels, respectively. (a1) and (b1): $(k_{\rm F}a_s)^{-1}=-1$ (BCS side). (a2) and (b2): $(k_{\rm F}a_s)^{-1}=0.02$ ($\simeq$unitarity). (a3) and (b3): $(k_{\rm F}a_s)^{-1}=1.24$ (BEC side). The intensity is normalized by the inverse Fermi energy $\varepsilon_{\rm F}^{-1}$. This normalization is also used in Figs. \ref{fig11} and \ref{fig13}.}
\label{fig10}
\end{figure}

Indeed, Fig. \ref{fig10}(b1) shows that the single-particle spectral weight $A_{\rm c}({\bm p},\omega<0)$ has very weak but non-zero intensity in the BCS regime ($(k_{\rm F}a_s)^{-1}=-1$), which contributes to $\rho_{\rm c}(\omega<0)$ in Eq. (\ref{eq.24}), as well as $N_{\rm c}$ in Eq. (\ref{eq.26}). Since the spectral weight vanishes in the negative energy region in a free Fermi gas with $\mu_{\rm c}<0$, the non-zero spectral intensity $A_{\rm c}({\bm p},\omega<0)$ seen in Fig. \ref{fig10}(b1)-(b3) originates from pairing fluctuations associated with the OFR-induced pairing interaction.
\par
%%%%%%%%%%%%%%%%%%%%%%%%%%%%%%%%%%%%%%%%%%%%%%%%%%%%%%%%%%%%%%%%%%%%%%%%%%%%%
%%%%%%%%%%%%%%%%%%%%%%%%%%%%%%%%%%%%%%%%%%%%%%%%%%%%%%%%%%%%%%%%%%%%%%%%%%%%%
\par
%%%%%%%%%%%%%%%%%%%%%%%%%%%%%%%%%%%%%%%%%%%%%%%%%%%%%%%%%%%%%%%%%%%%%%%%%%%%%
%%%%%%%%%%%%%%%%%%%%%%%%%%%%%%%%%%%%%%%%%%%%%%%%%%%%%%%%%%%%%%%%%%%%%%%%%%%%%
\par
To understand background physics of strong-coupling corrections to single-particle quantities $\rho_\alpha(\omega)$ and $A_\alpha({\bm p},\omega)$ shown in Figs. \ref{fig7} and \ref{fig10}, it is convenient to treat pairing fluctuations in the static approximation\cite{Levin2005,Tsuchiya,Chen}. This approximation assumes that $\Gamma_{\alpha\alpha}({\bm q},i\nu_n)$ ($\alpha={\rm o,c}$) is enhanced in the low-energy and low-momentum region near $T_{\rm c}$, reflecting the development of fluctuations in the Cooper-channel. (Note that $\lambda_\pm({\bm q}=0,i\nu_0=0)$ in Eq. (\ref{eq.15}) diverges, when the Thouless criterion in Eq. (\ref{eq.19}) is satisfied at $T_{\rm c}$, so that the ATMA particle-particle scattering matrix in the open channel ${\tilde \Gamma}_{oo}({\bm q}=0,i\nu_n=0)$ in Eq. (\ref{eq.17}), as well as that in the closed channel ${\tilde \Gamma}_{\rm cc}({\bm q}=0,i\nu_n=0)$, also diverge at $T_{\rm c}$.) Using this, we approximate the ATMA self-energy ${\tilde \Sigma}({\bm p},i\omega_n)$ in Eq. (\ref{eq.18}) to\cite{Tsuchiya},
\begin{eqnarray}
{\tilde \Sigma}({\bm p},i\omega_n)
&\simeq&
G_\alpha^0(-{\bm p},-i\omega_n)\times 
T\sum_{{\bm q},\nu_n}{\tilde \Gamma}_{\alpha\alpha}({\bm q},i\nu_n)
\nonumber
\\
&=&
=-\Delta_{{\rm PG},\alpha}^2G_\alpha^0(-{\bm p},-i\omega_n),
\label{eq.27}
\end{eqnarray}
where $\Delta_{{\rm PG},\alpha}^2=-T\sum_{{\bm q},\nu_n}{\tilde \Gamma}_{\alpha\alpha}({\bm q},i\nu_n)$ is sometimes referred to as the pseudogap parameter\cite{Levin2005,Tsuchiya,Chen}, describing effects of pairing fluctuations in the static approximation. Substituting Eq. (\ref{eq.27}) into the Green's function in Eq. (\ref{eq.8}) (Note that TMA self-energy $\Sigma$ is replaced by ATMA self-energy ${\tilde \Sigma}$ in our theory.), we have
\begin{eqnarray}
G_\alpha({\bm p},i\omega_n)
&=&
{1 \over i\omega_n-\xi_{\bm p}^\alpha-
{\displaystyle \Delta_{{\rm PG},\alpha}^2 \over \displaystyle i\omega_n+\xi_{\bm p}^\alpha}
}
\nonumber
\\
&=&
-{i\omega_n+\xi_{\bm p}^\alpha \over \omega_n^2+{\xi_{\bm p}^\alpha}^2+\Delta_{{\rm PG},\alpha}^2
}.
\label{eq.28}
\end{eqnarray}
Here, the second line is just the same form as the diagonal component of the ordinary BCS single-particle Green's function in the superfluid state\cite{Schrieffer}. Thus, the single-particle excitations have the same forms as the Bogoliubov single-particle dispersions,
\begin{equation}
E_{{\bm p},\pm}^\alpha
=\pm\sqrt{{\xi_{\bm p}^\alpha}^2+\Delta_{{\rm PG},\alpha}^2}.
\label{eq.29}
\end{equation}
The first line in Eq. (\ref{eq.28}) indicates that pairing fluctuations described by the pseudogap parameter $\Delta_{{\rm PG},\alpha}$ induce coupling between the particle branch ($\omega_{\rm particle}=\xi_{\bm p}^\alpha$) and the hole branch ($\omega_{\rm hole}=-\xi_{\bm p}^\alpha$) in each open and closed channel. 
\par
In the static approximation, when we consider the weak-coupling regime of the open channel, the particle dispersion $\omega_{\rm particle}=\xi_{\bm p}^{\rm o}=\varepsilon_{\bm p}-\mu$ and the hole dispersion $\omega_{\rm hole}=-\xi_{\bm p}^{\rm o}=-\varepsilon_{\bm p}+\mu$ cross at $p=\sqrt{2m\mu}$, because of $\mu(T_{\rm c})>0$ (see Fig. \ref{fig5}(b)). Then, the particle-hole coupling described by $\Delta_{{\rm PG},\alpha}$ causes the level repulsion, leading to the opening of the (pseudo)gap $\Delta E^{\rm o}_{\rm G}=2\Delta_{\rm PG}^{\rm o}$ at $p=\sqrt{2m\mu}$ (which equals $E_{{\bm p},+}^{\rm o}-E_{{\bm p},-}^{\rm o}$ at this momentum). Indeed, one sees two spectral peak lines along $\omega_{\rm particle}$ and $\omega_{\rm hole}$, as well as the pseudogap structure around $p/k_{\rm F}=1$ and $\omega=0$ in Fig. \ref{fig10}(a1).
\par
The pseudogap develops with increasing the interaction strength, reflecting the increase of the pseudogap parameter $\Delta_{\rm PG}^{\rm o}$, as shown in Fig. \ref{fig10}(a2). At the same time, the chemical potential gradually deviates from the Fermi energy $\varepsilon_{\rm F}$, to be negative in the strong-coupling regime (see Fig. \ref{fig5}(b)). When the chemical potential is negative, the particle branch ($\omega_{\rm particle}=\varepsilon_{\bm p}+|\mu|$) no longer crosses the hole branch $\omega_{\rm hole}=-\xi_{\bm p}^{\rm o}=-\varepsilon_{\bm p}-|\mu|$. The ``Bogoliubov" dispersion $E^{\rm o}_{{\bm p},+}$ ($E^{\rm o}_{{\bm p},-}$) in Eq. (\ref{eq.29}) then monotonically increases (decreases) with increasing the momentum $p$. The spectral structure in Fig. \ref{fig10}(a3) is found to really reflect this feature. In the static approximation, the pseudogap size in the BEC regime (where $\mu<0$) is given by the minimum of the energy difference $E^{\rm o}_{{\bm p},+}-E^{\rm o}_{{\bm p},-}$ at ${\bm p}=0$, which equals, not $2\Delta_{\rm PG}^{\rm o}$, but 
\begin{equation}
\Delta E^{\rm o}_{\rm G}=2\sqrt{|\mu|^2+\Delta_{{\rm PG},{\rm o}}^2}.
\label{eq.30}
\end{equation}
\par
In the closed channel, the effective chemical potential $\mu_{\rm c}=\mu-\nu/2$ in the dispersion $\xi_{\bm p}^{\rm c}=\varepsilon_{\bm p}-\mu_{\rm c}$ is always {\it negative} in the whole BCS-BEC crossover region at $T_{\rm c}$ (see Fig. \ref{fig8}), which is similar to the strong-coupling case in the open channel (where $\mu<0$). Indeed, the overall spectral structures in Fig. \ref{fig10}(b1)-(b3) are similar to that shown in Fig. \ref{fig10}(a3). 
\par
However, while the pseudogap size $\Delta E_{\rm G}^{\rm o}$ in the open channel in Eq. (\ref{eq.30}) increases with increasing the interaction strength in the BEC regime (because of the increase of $|\mu|$), the opposite tendency is seen in the closed channel, as shown in Figs. \ref{fig10}(b1)-(b3). This is simply because the magnitude $|\mu_{\rm c}|$ of the effective chemical potential {\it decreases} with increasing the interaction strength (see Fig. \ref{fig8}), so that the pseudogap size in the closed channel,
\begin{equation}
\Delta E^{\rm c}_{\rm G}
=2\sqrt{|\mu_{\rm c}|^2+\Delta_{{\rm PG},{\rm o}}^2},
\label{eq.31}
\end{equation}
also decreases, as one passes through the BCS-BEC crossover region. The decrease of $|\mu_{\rm c}|=|\mu-\nu/2|$ dominantly originates from the decrease of the band gap $\nu/2$ in tuning the strength of the OFR-induced pairing interaction.
\par
%%%%%%%%%%%%%%%%%%%%%%%%%%%%%%%%%%%%%%%%%%%%%%%%%%%%%%%%%%%%%%%%%%%%%%%%%%%%%
\begin{figure}[t]
\center
\includegraphics[width=0.45\textwidth]{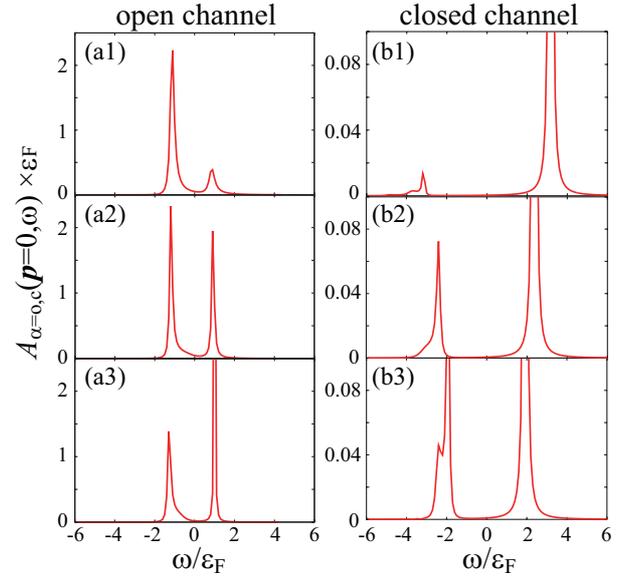}
\caption{(Color online) Single-particle spectral weight $A_{\alpha={\rm o,c}}({\bm p}=0,\omega)$, as a function of $\omega$. We take $T=T_{\rm c}$. The left (right) panels show the results in the open (closed) channel. (a1) and (b1): $(k_{\rm F}a_s)^{-1}=-0.6$ ($T_{\rm c}=0.212T_{\rm F}$). (a2) and (b2): $(k_{\rm F}a_s)^{-1}=0.02$ ($T_{\rm c}=0.248T_{\rm F}$).  (a3) and (b3): $(k_{\rm F}a_s)^{-1}=0.4$.($T_{\rm c}=0.246T_{\rm F}$).}
\label{fig11}
\end{figure}
%%%%%%%%%%%%%%%%%%%%%%%%%%%%%%%%%%%%%%%%%%%%%%%%%%%%%%%%%%%%%%%%%%%%%%%%%%%%%
\par
The importance of the different interaction dependence between $\mu$ and $\mu_{\rm c}$ also appears in considering the spectral intensity around the upper branch $\omega=E_{{\bm p},+}^\alpha~(>0)$ and that around the lower branch $E_{{\bm p},-}^\alpha~(<0)$. Substituting Eq. (\ref{eq.28}) into Eq. (\ref{eq.25}), one has
\begin{equation}
A_\alpha({\bm p},\omega)=
{1 \over 2}
\left[
1+{\xi_{\bm p}^\alpha \over E_{{\bm p},+}^\alpha}
\right]
\delta(\omega-E_{{\bm p},+}^\alpha)
+
{1 \over 2}
\left[
1+{\xi_{\bm p}^\alpha \over E_{{\bm p},-}^\alpha}
\right]
\delta(\omega-E_{{\bm p},-}^\alpha).
\label{eq.32}
\end{equation}
In particular at ${\bm p}=0$, Eq. (\ref{eq.32}) becomes
\begin{eqnarray}
A_\alpha(0,\omega)
=
F_+^\alpha\delta
\left(
\omega-\sqrt{\mu_\alpha^2+\Delta_{{\rm PG},\alpha}^2}
\right)
+
F_-^\alpha
\delta
\left(
\omega+\sqrt{\mu_\alpha^2+\Delta_{{\rm PG},\alpha}^2}
\right),
\label{eq.33}
\end{eqnarray}
where 
\begin{equation}
F_\pm^{\alpha={\rm o,c}}=
{1 \over 2}
\left[
1\mp {{\mu_\alpha \over \sqrt{\mu_\alpha^2+\Delta_{{\rm PG},\alpha}^2}}}
\right],
\label{eq.34}
\end{equation}
with $\mu_{\rm o}\equiv \mu$ and $\mu_{\rm c}=\mu-\nu/2$. In the open channel, as the interaction strength increases, the fact that $\mu_o~(=\mu)$ monotonically decreases leads to the increase of the spectral intensity $F_+^{\rm o}$ of the upper branch ($\omega=\sqrt{\mu_{\rm o}^2+\Delta_{{\rm PG},{\rm o}}^2}$), as well as the decrease of $F_-^{\rm o}$ of the lower branch ($\omega=-\sqrt{\mu_{\rm o}^2+\Delta_{{\rm PG},{\rm o}}^2}$). 
\par
In contrast, because $\mu_{\rm c}~(=\mu-\nu/2)$ in the closed channel monotonically increases with increasing the interaction strength, the spectral intensity $F_+^{\rm c}$ of the upper branch and the that ($F_-^{\rm c}$) of the lower branch, respectively, exhibit the opposite interaction dependence to the open-channel case. Although the above discussion is within the simple static approximation, the same conclusion is obtained without using this approximation, as shown in Fig. \ref{fig11}.
\par
%%%%%%%%%%%%%%%%%%%%%%%%%%%%%%%%%%%%%%%%%%%%%%%%%%%%%%%%%%%%%%%%%%%%%%%%%%%%%
\begin{figure}[t]
\center
\includegraphics[width=0.45\textwidth]{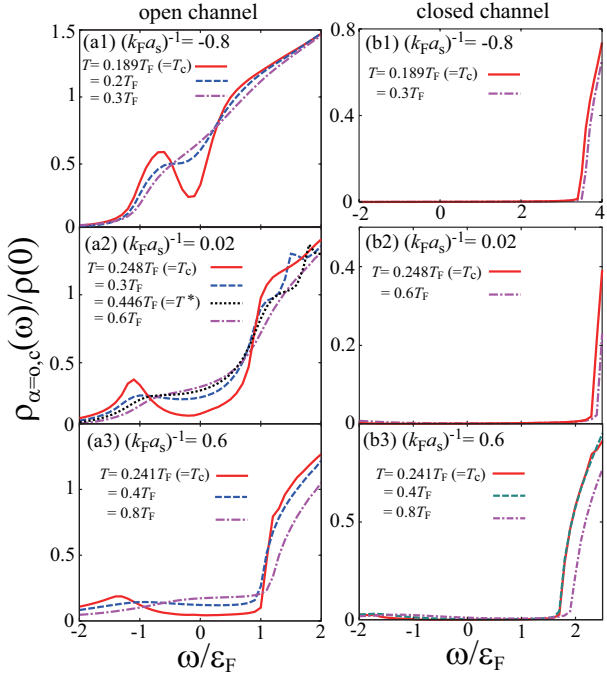}
\caption{(Color online) DOS $\rho_{\alpha={\rm o,c}}(\omega)$ at various temperatures above $T_{\rm c}$. In panel (a2), $T^*$ is the pseudogap temperature which is defined as the temperature above which a dip no longer exists in $\rho_{\rm o}(\omega\sim 0)$. When $(k_{\rm F}a_s)^{-1}=-0.8$ (panel (b1)) and $0.02$ (panel (b2)), since the temperature dependence of $\rho_{\rm c}(\omega)$ is very weak, we only show the results at $T_{\rm c}$ and at a high temperature. We briefly note that the peak structures seen around $\omega/\varepsilon_{\rm F}=1.5$ in panel (a2) are due to numerical problems in carrying out the analytic continuation by the Pad\'e approximation\cite{Vidberg}.}
\label{fig12}
\end{figure}
%%%%%%%%%%%%%%%%%%%%%%%%%%%%%%%%%%%%%%%%%%%%%%%%%%%%%%%%%%%%%%%%%%%%%%%%%%%%%
\par
\begin{figure}[t]
\center
\includegraphics[width=0.45\textwidth]{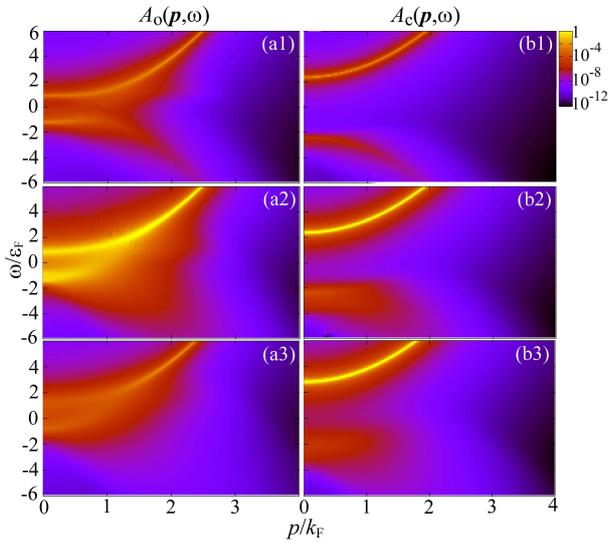}
\caption{(Color online) Intensity of single-particle spectral weight $A_{\alpha={\rm o,c}}({\bm p},\omega)$ above $T_{\rm c}$. We take $(k_{\rm F}a_s)^{-1}=0.02$. The left and right panels show the results in the open channel and closed channel, respectively. (a1) and (b1): $T=T_{\rm c}=0.248T_{\rm F}$. (a2) and (b2): $T=0.5T_{\rm F}$. (a3) and (b3): $T=0.9T_{\rm F}$.}
\label{fig13}
\end{figure}

Figure \ref{fig12} shows DOS $\rho_\alpha(\omega)$ above $T_{\rm c}$. In the open channel shown in panels (a1)-(a3), the temperature dependence of the pseudogap is qualitatively the same as that in the single-channel case discussed in the BCS-BEC crossover regime of $^{40}$K and $^6$Li Fermi gases\cite{Tsuchiya}: The pseudogap gradually disappears with increasing the temperature, reflecting the weakening of pairing fluctuations. Correspondingly, two spectral peak lines along the particle and hole dispersions in the spectral weight $A_{\rm o}({\bm p},\omega)$ becomes obscure at high temperatures, as shown in Figs. \ref{fig13}(a1)-(a3). As in the single-channel case\cite{Tsuchiya,Tsuchiya2}, introducing the pseudogap temperature $T^*$ as the temperature at which a dip around $\omega=0$ disappears in DOS $\rho_{\rm o}(\omega)$, one finds, for example, $T^*/T_{\rm F}\simeq 0.45$ in the unitarity limit. This value is somehow higher than the single-channel case, $T^*/T_{\rm F}\simeq 0.3$, at this interaction strength\cite{Tsuchiya}. These results indicate that, as in $^{40}$K and $^6$Li Fermi gases, we can also examine the pseudogap phenomenon in the open channel of a $^{173}$Yb Fermi gas, when the interaction strength is tuned by OFR.
\par
%%%%%%%%%%%%%%%%%%%%%%%%%%%%%%%%%%%%%%%%%%%%%%%%%%%%%%%%%%%%%%%%%%%%%%%%%%%%%
%%%%%%%%%%%%%%%%%%%%%%%%%%%%%%%%%%%%%%%%%%%%%%%%%%%%%%%%%%%%%%%%%%%%%%%%%%%%%
\par
On the other hand, Figs. \ref{fig12}(b1)-(b3) show that DOS in the closed channel is not sensitive to the temperature. This is simply because of the existence of the large band gap $\nu/2$ between the open and closed channels. Although this gap becomes small in the strong-coupling BEC regime, this regime is then dominated by two-body bound molecules with a large binding energy $E_{\rm bind}$, so that we again do not expect remarkable temperature dependence of $\rho_{\rm c}(\omega)$ there (unless we consider the high temperature region, $T\gesim E_{\rm bind}$). 
\par
As expected from Figs. \ref{fig12}(b1)-(b3), the spectral weight $A_{\rm c}({\bm p},\omega)$ in the closed channel is also not sensitive to the temperature compared to the open channel case, as shown in Figs. \ref{fig13}(b1)-(b3). Because of this $T$-insensitive spectral weight $A_{\rm c}({\bm p},\omega)$ in the closed channel, as well as the existence of a relatively large band gap $\nu/2$ between the open and closed channel (except near $(k_{\rm F}a_s)^{-1}=1.57$), the number $N_{\rm c}$ of atoms in the closed channel is not sensitive to the temperature, as shown in Fig. \ref{fig9}(b).
\par
Figure \ref{fig13}(b3) indicates that, in the unitarity regime, the lower spectral peak in $A_{\rm c}({\bm p},\omega)$ remains to exist to $T/T_{\rm F}\sim 0.9$. Since the appearance of this lower branch is a strong-coupling phenomenon associated with the OFR-induced pairing interaction, it would be an interesting challenge to observe it by using the photoemission-type experiment\cite{Tsuchiya2,Torma,Stewart,Gaebler,Sagi}. Regarding this, we recall that the photoemission-type experiment detects, roughly speaking, the product of the spectral weight $A_\alpha({\bm p},\omega)$ and the Fermi distribution function $f(\omega)$, so that the spectral intensity in the negative energy region tends to be emphasized. In this sense, this experimental technique would be suitable for our purpose.
\par
%%%%%%%%%%%%%%%%%%%%%%%%%%%%%%%%%%%%%%%%%%%%%%%%%%%%%%%%%%%%%%%%%%%%%%%%%%%%%%
\par
\section{Summary}
\par
To summarize, we have discussed single-particle excitations and strong-coupling effects in an ultracold Fermi gas with an orbital Feshbach resonance (OFR). In particular, we have picked up a $^{173}$Yb Fermi gas, where OFR has recently been observed. Including strong-pairing fluctuations within the framework of a $T$-matrix approximation (TMA), we have calculated the single-particle density of states (DOS) $\rho_\alpha(\omega)$, as well as the spectral weight $A_\alpha({\bm p},\omega)$, in both the open ($\alpha={\rm o}$) and closed ($\alpha={\rm c}$) channels, in the normal state above $T_{\rm c}$. In constructing the theory, we explained how to remove unwanted effects coming from an experimentally inaccessible deep bound state.
\par
We clarified how single-particle properties vary, when one tunes the strength of an OFR-induced pairing interaction by adjusting the band-gap $\nu/2$ between the open and closed channels. In the open channel near $T_{\rm c}$, strong pairing fluctuations were shown to cause the pseudogap phenomenon, where a dip appears in DOS $\rho_{\rm o}(\omega)$ around $\omega=0$, which grows as one passes through the BCS-BEC crossover region, to become a large gap in the strong-coupling BEC regime. Correspondingly, the single-particle spectral weight $A_{\rm o}({\bm p},\omega)$ in the BCS side exhibits a coupling phenomenon between the particle and hole excitations by pairing fluctuations, giving the pseudogap around $\omega=0$. In the strong-coupling regime where the chemical potential $\mu$ is negative, the particle-hole coupling is no longer seen in $A_{\rm o}({\bm p},\omega)$; the particle excitations and hole excitations separately produce the spectral intensity in the positive and negative energy region, respectively. With increasing the temperature, these pseudogap phenomena gradually become obscure, due to the suppression of pairing fluctuations. Apart from details, these results in the open channel are qualitatively the same as the BCS-BEC crossover behaviors of single-particle excitations discussed in the single-channel case with a broad magnetic Feshbach resonance describing $^{40}$K and $^6$Li Fermi gases. Thus, the study of the open channel in a $^{173}$Yb would provide useful opportunity to confirm to what extent the BCS-BEC crossover discussed in alkali-metal Fermi gases is a universal phenomenon.
\par
In contrast, the interaction dependence of DOS $\rho_{\rm c}(\omega)$ in the close channel is very different from that in the open channel. In the BCS side ($(k_{\rm F}a_s)^{-1}\lesssim 0$), one cannot see a remarkable strong-coupling correction to $\rho_{\rm c}(\omega)$, because a large band gap $\nu/2$ suppresses the low-energy part of this quantity. In the BEC side ($(k_{\rm F}a_s)^{-1}\gesim 0$), on the other hand, the small band gap $\nu/2$, as well as strong pairing interaction, lead to the increase of $\rho_{\rm c}(\omega)$ in the negative energy region. The band gap between the open and closed channels vanishes at $k_{\rm F}a_s)^{-1}=1.57$ in the BEC regime, where $\rho_{\rm c}(\omega)=\rho_{\rm o}(\omega)$ is realized.
\par
However, even in the weak-coupling BCS side, the single-particle spectral weight $A_{\rm c}({\bm p},\omega)$ in the closed channel possesses weak but non-zero spectral intensity in the negative energy region, in addition to the dominant spectral intensity above the band gap energy $\omega\gesim \nu/2$. (Although the spectral intensity in the negative energy region should contribute to $\rho_{\rm c}(\omega<0)$, it is actually very small in the BCS regime.) While the spectral intensity in the negative energy region increases with increasing the interaction strength, the spectral intensity in the positive energy region decreases. This result is opposite to the open channel case, where the spectral intensity in the negative (positive) energy region decreases (increases) as the interaction strength increases. We pointed out that this difference between the two channels originate from the different interaction dependence of the chemical potential $\mu$ in the open channel and the effective chemical potential $\mu_{\rm c}=\mu-\nu/2$ in the closed channel. 
\par
In a broad magnetic Feshbach resonance (MFR) used in alkali metal $^{40}$K and $^6$Li Fermi gases, the closed channel only virtually contributes to the pairing interaction. As a result, the BCS-BEC crossover phenomenon has only been examined in the open channel. On the other hand, our results indicate that a rare-earth $^{173}$Yb Fermi gas with an orbital Feshbach resonance provides useful opportunity to examine, not only the BCS-BEC crossover behaviors of the open channel, but also strong-coupling effects on normal-state properties of the closed channel. The observation of the latter is an advantage of a $^{173}$Yb Fermi gas with OFR. 
\par
Regarding this, the photoemission-type experiment would be useful, because this experimental technique can observe the single-particle spectral weight $A_{\rm c}({\bm p},\omega)$ multiplied by the Fermi distribution function $f(\omega)$, so that the spectral intensity in the negative region is emphasized. Since the observed photoemission spectrum would be affected by spatial inhomogeneity associated with a harmonic trap, to theoretically evaluate this quantity, we need to extend our analyses for a uniform Fermi gas to include the trapped geometry, which remains as our future problem. Since OFR is expected as a promising pairing mechanism to realize the superfluid phase transition in a $^{173}$Yb Fermi gas, our results would be useful in considering the similarity and difference between this rare-earth Fermi gas and alkali metal $^{40}$K and $^6$Li Fermi gases. 
\par
%%%%%%%%%%%%%%%%%%%%%%%%%%%%%%%%%%%%%%%%%%%%%%%%%%%%%%%%%%%%%%%%%%%%%%%%%%%%%%
\par
\vskip2mm
\noindent
{\bf Acknowledgments}
\par
We thank R. Hanai, D. Kharga, P. van Wyk, M. Matsumoto, and D. Kagamihara for useful discussions. This work was supported by KiPAS project in Keio University. D.I. was supported by Grant-in-aid for Scientific Research from JSPS in Japan (No.JP16K17773). Y.O. was supported by Grant-in-aid for Scientific Research from JSPS in Japan (No.JP15H00840, No.JP15K00178, No.JP16K05503)
\par
%%%%%%%%%%%%%%%%%%%%%%%%%%%%%%%%%%%%%%%%%%%%%%%%%%%%%%%%%%%%%%%%%%%%%%%%%%%%%%

%%%%%%%%%%%%%%%%%%%%%%%%%%%%%%%%%%%%%%%%%%%%%%%%%%%%%%%%%%%%%%%%%%%%%%%%%%%%%%%

\begin{thebibliography}{99}
\bibitem{note} For simplicity, we ignore other interactions, such as an interaction between $|{\rm e}\rangle_{\rm e}||a\rangle_{\rm n}$ and $|{\rm e}\rangle_{\rm e}||b\rangle_{\rm n}$. The residual interactions are usually weak, compared to the OFR-induced one in the BCS-BEC crossover region.
\end{thebibliography}
\end{document}